\documentclass[10pt,preprint]{emulateapj}  

\usepackage{color}
\usepackage{amsmath}
\usepackage{epsfig}
\usepackage{subfigure,verbatim}
\usepackage[colorlinks,urlcolor=blue,citecolor=blue,linkcolor=blue]{hyperref}

\newcommand{\bmath}[1]{\mbox{\boldmath{$#1$}}}

\newcommand{\bi}{\begin{itemize}}
\newcommand{\ei}{\end{itemize}}

\def\comp{\,c/\omega_{\rm p}}

\def\ompt{\omega_{\rm p}t}

\newcommand{\fig}[1]{Fig.~\ref{fig:#1}}
\newcommand{\be}{\begin{eqnarray}}
\newcommand{\ee}{\end{eqnarray}}

\begin{document}
\title{Relativistic Reconnection: an Efficient Source of Non-Thermal Particles}
\author{Lorenzo Sironi$^{1,2}$ and Anatoly Spitkovsky$^3$}
\affil{$^1$Harvard-Smithsonian Center for Astrophysics, 
60 Garden Street, Cambridge, MA 02138, USA; lsironi@cfa.harvard.edu
\\
$^2$NASA Einstein Postdoctoral Fellow
\\
$^3$Department of Astrophysical Sciences, Princeton University, Princeton, NJ 08544-1001, USA; anatoly@astro.princeton.edu}
 
\begin{abstract}
In magnetized astrophysical outflows, the dissipation of field energy into particle energy via magnetic reconnection is often invoked to explain the observed non-thermal signatures. By means of two- and three-dimensional particle-in-cell simulations, we investigate anti-parallel reconnection in magnetically-dominated electron-positron plasmas. Our simulations extend to unprecedentedly long temporal and spatial scales, so we can capture the asymptotic state of the system beyond the initial transients, and without any artificial limitation by the boundary conditions. At late times, the reconnection layer is organized into a chain of large magnetic islands connected by thin X-lines. The plasmoid instability further fragments each X-line into a series of smaller islands, separated by X-points.
At the X-points, the particles become unmagnetized and they get accelerated along the reconnection electric field. We provide definitive evidence that the late-time particle spectrum integrated over the whole reconnection region is a power-law, whose slope is harder than $-2$ for magnetizations $\sigma\gtrsim 10$. Efficient particle acceleration to non-thermal energies is a generic by-product of the long-term evolution of relativistic reconnection in both two and three dimensions. In three dimensions, the drift-kink mode corrugates the reconnection layer at early times, but the long-term evolution is controlled by the plasmoid instability, that facilitates efficient particle acceleration, in analogy to the two-dimensional physics. Our findings have important implications for the generation of hard photon spectra in pulsar winds and relativistic astrophysical jets.
\end{abstract}

\keywords{acceleration of particles --- galaxies: jets --- gamma-ray burst: general --- magnetic reconnection --- pulsars: general --- radiation mechanisms: non-thermal}

\section{Introduction}
It is generally thought that pulsar winds and the relativistic jets of blazars and gamma-ray bursts are launched hydromagnetically \citep[][]{spruit_10}. Since the energy is initially carried  in the form of Poynting flux, it is a fundamental question how the field energy is transferred to the plasma, to power the observed emission. Field dissipation via magnetic reconnection is often invoked as a source of the accelerated particles required to explain the non-thermal signatures of pulsar wind nebulae \citep[PWNe;][]{lyubarsky_kirk_01,lyubarsky_03,kirk_sk_03,petri_lyubarsky_07}, jets from active galactic nuclei \citep[AGNs;][]{romanova_92,giannios_09,giannios_13} and gamma-ray bursts \citep[GRBs;][]{thompson_94, thompson_06,spruit_01,lyutikov_03,giannios_08}. Despite decades of research, the efficiency of magnetic reconnection in generating non-thermal particles is not well understood \citep[][]{hoshino_lyubarsky_12}.

In astrophysical jets, reconnection proceeds in the ``relativistic'' regime, since the magnetic energy per particle can exceed the rest mass energy. While the steady-state dynamics of relativistic  reconnection has been well characterized by analytical studies \citep[][]{lyutikov_uzdensky_03,lyubarsky_05}, the process of particle acceleration can only be captured from first principles by means of fully-kinetic particle-in-cell (PIC) simulations. Energization of particles in relativistic reconnection has been investigated in a number of PIC studies, both in two dimensions \citep[2D;][]{zenitani_01,zenitani_07,jaroschek_04,bessho_05,bessho_07,bessho_12,daughton_07,lyubarsky_liverts_08} and three dimensions \citep[3D;][]{zenitani_08,yin_08,liu_11,sironi_spitkovsky_11b, sironi_spitkovsky_12,kagan_13,cerutti_13b}. Yet, no consensus exists whether relativistic reconnection results self-consistently in non-thermal particle acceleration \citep[][]{sironi_spitkovsky_11b}, rather than just heating \citep[][]{cerutti_12a}.

\begin{figure*}[!tbp]
\begin{center}
\includegraphics[width=1.05\textwidth]{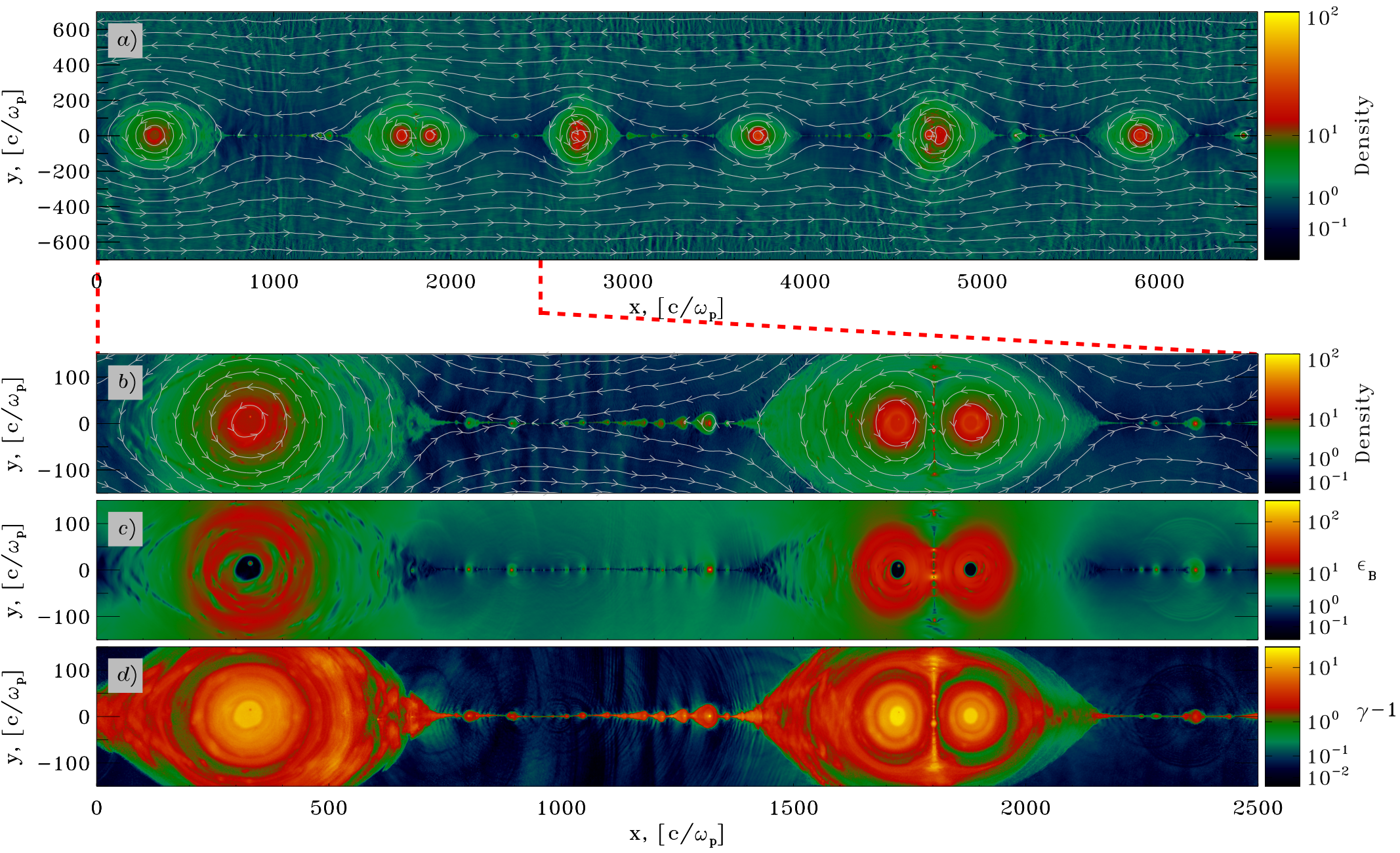}
\caption{Structure of the reconnection layer at $\ompt=3000$ (so, $\omega_c t\sim 10^4$), from a 2D simulation of $\sigma=10$ reconnection. The box extends along $x$ over $\sim 6550\comp$ (65536 cells), and along $y$ over $\sim 6000\comp$ ($\sim60000$ cells), but along $y$ it will expand even more at later times (we only show a subset of the domain along $y$). We present (a) particle density, in units of the density far from the  sheet (with overplotted magnetic field lines), (b) magnetic energy fraction $\epsilon_B=B^2/8\pi m n c^2$ and (c) mean kinetic energy per particle.}
\label{fig:fluid2d}
\end{center}
\end{figure*}

In this work, we employ 2D and 3D PIC simulations to follow the evolution of relativistic reconnection in pair plasmas to unprecedentedly long time and length scales, focusing on particle acceleration. We consider the case of anti-parallel fields, without a guide field aligned with the electric current in the sheet. It has been argued that this configuration produces non-thermal particles only in 2D, whereas in 3D the drift-kink (DK) mode would broaden the current sheet, inhibiting particle acceleration \citep{zenitani_08,cerutti_13b}. By performing large-scale simulations evolved to long times, we conclusively show that acceleration of particles to non-thermal energies is a {\it generic} by-product of relativistic reconnection in pair plasmas, in both 2D and 3D. The accelerated particles populate a power-law distribution, whose spectral slope is harder than $-2$ for magnetizations $\sigma\gtrsim10$. Relativistic magnetic reconnection is then a viable source of non-thermal emission from magnetically-dominated astrophysical flows. 

\begin{figure}[!tbp]
\begin{center}
\includegraphics[width=0.51\textwidth]{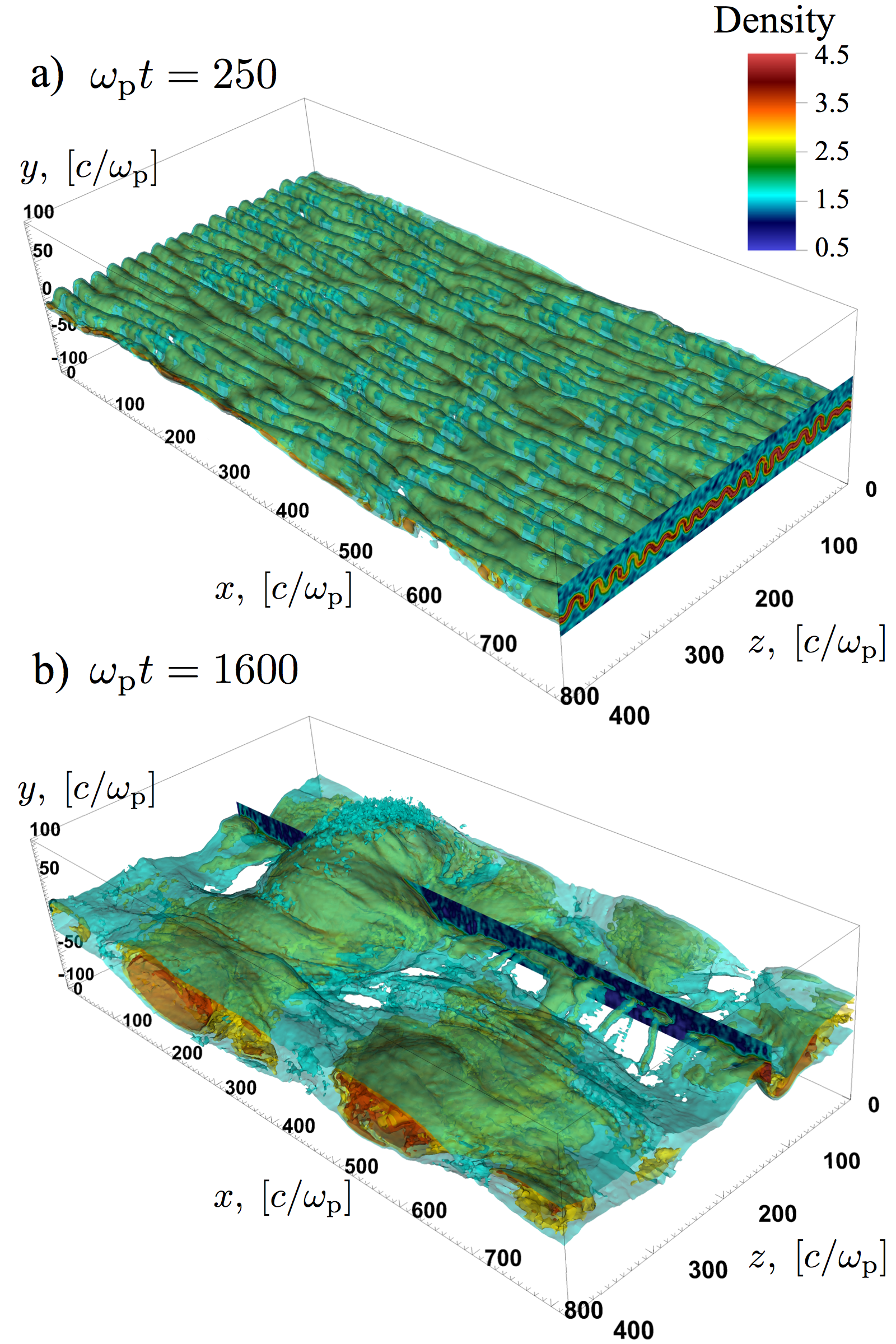}
\caption{Structure of particle density at (a) $\ompt=250$ and (b) $\ompt=1600$, from a 3D simulation of $\sigma=10$ reconnection without guide field. The simulation box extends over $\sim 820\comp$ (4096 cells) along $x$, $\sim 410\comp$ (2048 cells) along $z$, and it expands at the speed of light along $\pm \hat{\bmath{y}}$ (we only show a subset of the domain along $y$). The 2D slices in the top and bottom panels (at $x=820\comp$ and $z=130\comp$, respectively) show the particle density.}
\label{fig:fluid3d}
\end{center}
\end{figure}

\section{Structure of the Reconnection Layer}\label{sec:struct}
We use the 3D electromagnetic PIC code TRISTAN-MP \citep{buneman_93, spitkovsky_05} to study relativistic reconnection in 2D and 3D. The reconnection layer is set up in Harris equilibrium, with the magnetic field $\bmath{B}=-B_0\, \bmath{\hat{x}}\tanh(2\pi y/\Delta)$  reversing at $y=0$. The field strength is parameterized by the magnetization $\sigma=B_0^2/4\pi m n c^2=(\omega_c/\omega_{\rm p})^2$, where $\omega_c=eB_0/mc$ is the Larmor frequency and $\omega_{\rm p}=\sqrt{4\pi n e^2/m}$ is the plasma frequency for the electron-positron plasma outside the layer. We focus on the regime $\sigma\geq1$ of relativistic reconnection. The magnetic pressure outside the current sheet is balanced by the particle pressure in the sheet, by adding a component of hot plasma with overdensity $\eta$ relative to the cold particles outside the layer (having $k_B T/m c^2=10^{-4}$). From pressure equilibrium, the temperature inside the sheet is $k_B T_{h}/m c^2=\sigma/2\eta$. We typically employ  $\eta=3$ and $\Delta=20\comp$ ($c/\omega_{\rm p}$ is the plasma skin depth), but we have tested that our results at late times do not depend on the initialization of the current sheet (Sironi 2014, in preparation; hearafter S14).

In 2D, the computational domain is periodic in the $x$ direction (in 3D, in $x$ and $z$), but we have extensively tested that the results reported in this work are not artificially affected by our periodic boundaries (which is often an issue for smaller simulations, S14). Along the $y$ direction, we employ two Òmoving injectorsÓ (receding from $y=0$ at the speed of light along $\pm \hat{\bmath{y}}$) and an expanding simulation box  (S14). 
The two injectors constantly introduce fresh magnetized  plasma into the simulation domain. This permits us to evolve the system as far as the computational resources allow, retaining all the regions that are in causal contact with the initial setup. Such choice has clear advantages over the fully-periodic setup that is commonly employed, where the limited amount of particles and magnetic energy will necessarily inhibit the evolution of the system to long times. 

For our reference case $\sigma=10$, we resolve the plasma skin depth with $\comp=10$  cells in 2D and 5 cells in 3D, and for higher magnetizations we scale up the resolution by $\sqrt{\sigma/10}$, so that the Larmor gyration period $2\pi/\omega_c=2\pi/\sqrt{\sigma}\,\omega_{\rm p}$ is resolved with at least a few timesteps. We typically employ four particles per cell in 2D and one per cell in 3D (on average, including both species), but we have extensively tested that the  physics at late times is the same when using up to 64 (in 2D) or 8 (in 3D) particles per cell (S14). 

Magnetic reconnection starts from numerical noise (unlike most other studies, we do not artificially perturb the magnetic flux function to trigger reconnection). As a result of the tearing instability, the reconnection layer breaks into a series of magnetic islands, separated by X-points. Over time, the islands coalesce and grow to larger scales (\citealt{daughton_07}, for similar conclusions in non-relativistic reconnection). The structure of the reconnection region at late times is presented in \fig{fluid2d}, from our large-scale 2D simulation in a $\sigma=10$ pair plasma. The reconnection layer at $\ompt=3000$ is divided into six major islands, separated by thin X-lines (\fig{fluid2d}a). By zooming into the region $x\lesssim 2500\comp$, as indicated by the dashed red lines below \fig{fluid2d}a, we show that each X-line is further fragmented into a chain of smaller islands, as a result of the secondary tearing mode (or ``plasmoid instability'') discussed by \citet{uzdensky_10}. The secondary islands appear as  overdense regions at $700\comp\lesssim x \lesssim 1400\comp$ in \fig{fluid2d}b, which are filled with hot particles (\fig{fluid2d}d) confined by strong fields (\fig{fluid2d}c). In between each pair of secondary islands, a secondary X-point governs the transfer of energy from the fields to the particles. As shown in \S\ref{sec:accel}, efficient particle acceleration occurs at the X-points (both at secondary X-points and at the primary X-point located at $x\sim1000\comp$ in \fig{fluid2d}b-d).

The cold upstream plasma flows into the X-line at the speed $v_{\rm rec}\simeq 0.08\,c$ (the so-called ``reconnection rate''), nearly constant over time. As shown in S14, the reconnection rate increases from $v_{\rm rec}\simeq 0.03\,c$ for $\sigma=1$ to $v_{\rm rec}\simeq 0.12\,c$ for $\sigma=30$, and it is nearly independent of $\sigma$ for larger magnetizations (we have tried up to $\sigma=100$), in agreement with \citet{lyubarsky_05}. After entering the current sheet, the flow is advected towards the  major islands by the reconnected magnetic field (in the inset of \fig{fluid2d}b-d, the major islands lie at $200\comp\lesssim x\lesssim500\comp$ and $1600\comp\lesssim x\lesssim1900\comp$). The fast reconnection exhausts move with a bulk Lorentz factor $\sim \sqrt{\sigma}\simeq 3$, as predicted by \citet{lyubarsky_05}. Eventually, all the particles heated and accelerated by reconnection are trapped within the major islands, which act as reservoirs of particles (\fig{fluid2d}b) and particle energy (\fig{fluid2d}d). There, the magnetic field is also stronger (by effect of compression, \fig{fluid2d}c), so we expect the major islands to dominate the synchrotron emissivity.

Pushed by the ram pressure of the reconnection exhausts, the major islands move along the  layer, merging with neighboring islands. Two major islands are coalescing at $x\sim 1800\comp$ in \fig{fluid2d}. The current sheet formed between the two merging islands is tearing-unstable, and it breaks into a chain of secondary islands  along the $y$ direction. As shown in S14, the anti-reconnection electric field at the X-points between two merging islands plays an important role in boosting the high-energy particles trapped in the islands to even higher energies.\footnote{The hierarchical process of island formation and merging proceeds as long as the evolution is not artificially inhibited by the boundary conditions. This emphasizes the importance of large-scale simulations to properly capture the late-time physics.}

The evolution of 3D reconnection at late times parallels closely the 2D physics described above,  {\it even in the absence of a guide field}. As shown in \fig{fluid3d}a, we find that the early phases are governed by the DK mode, in agreement with previous studies \citep{zenitani_08,cerutti_13b}. The DK instability corrugates the current sheet in the $z$ direction, broadening the  layer and inhibiting the early growth of the tearing mode. However, at later times the evolution is controlled by the tearing instability, that produces in the $xy$ plane a sequence of magnetic islands (or rather, tubes) that  grow and coalesce over time, in analogy to the 2D physics. The reconnection layer at late times is organized into a few major islands (see the overdense plasmoids in \fig{fluid3d}b), separated by underdense regions (transparent in \fig{fluid3d}b) where field dissipation by reconnection is most efficient. In such neutral planes, the secondary tearing instability can operate, producing a chain of smaller magnetic tubes (see the region at $500\comp\lesssim x\lesssim 700\comp$ near the 2D slice at $z=130\comp$ in \fig{fluid3d}b), reminiscent of the secondary islands observed in 2D. In short, at late times the 3D physics parallels closely the 2D evolution presented above (albeit with a reduced reconnection rate, $v_{\rm rec}\simeq0.02\,c$ in 3D versus $v_{\rm rec}\simeq0.08\,c$ in 2D). As shown in \S\ref{sec:accel}, this has important implications for the acceleration performance of relativistic reconnection in 3D. Earlier simulations, being limited to early times and/or small boxes, could not capture the long-term evolution of 3D reconnection, with the emergence of the dominant tearing mode beyond the DK phase.

\begin{figure}[tbp]
\begin{center}
\includegraphics[width=0.51\textwidth]{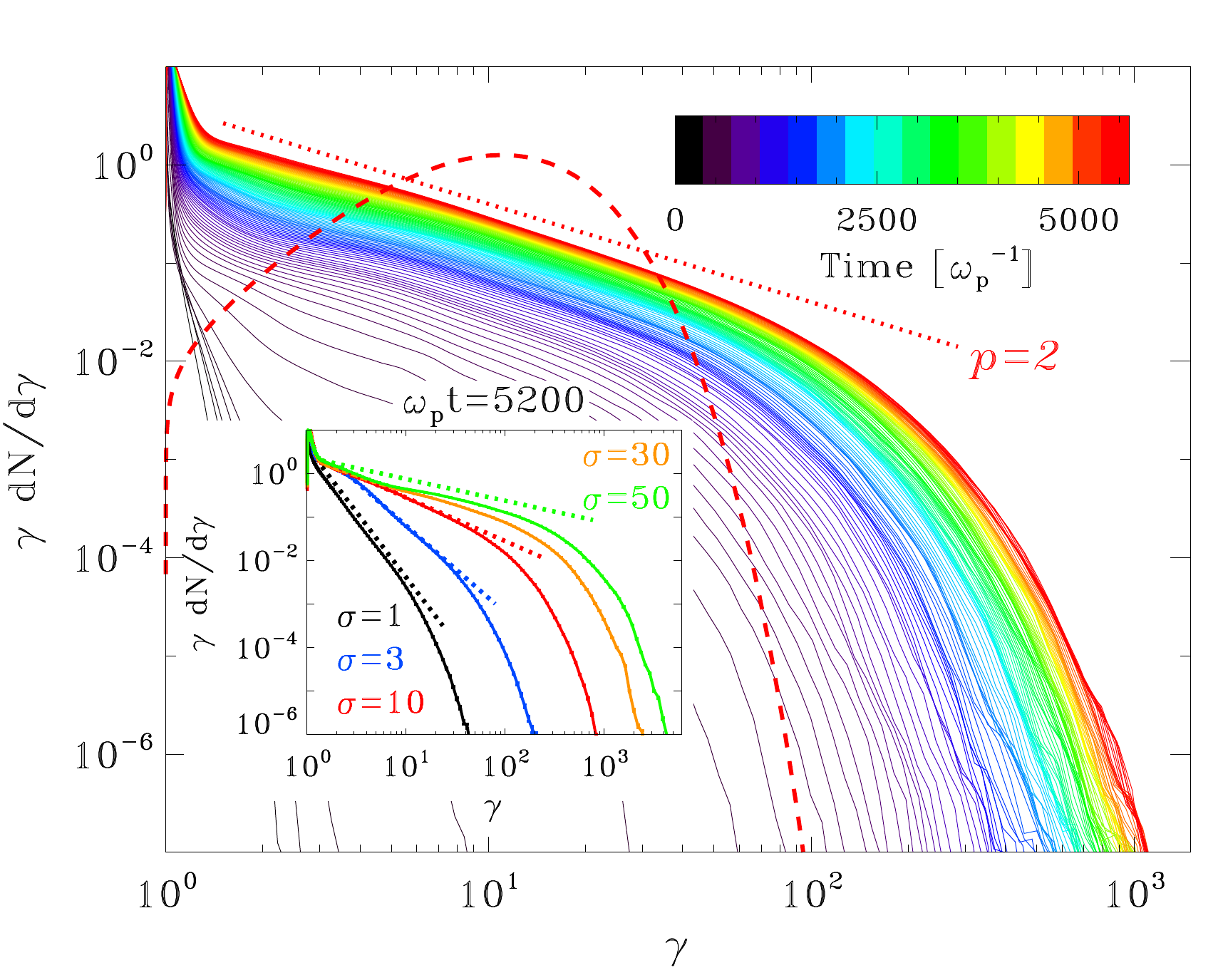}
\caption{Temporal evolution of particle energy spectrum, from a 2D simulation of $\sigma=10$ reconnection. The spectrum at late times resembles a power-law with slope $p=2$ (dotted red line), and it clearly departs from a Maxwellian with mean energy $(\sigma+1)\,mc^2$ (dashed red line, assuming complete field dissipation). In the inset, dependence of the spectrum on the magnetization, as indicated in the legend. The dotted lines refer to power-law slopes of $-4$, $-3$, $-2$ and $-1.5$ (from black to green).}
\label{fig:spec2d}
\end{center}
\end{figure}

\begin{figure}[tbp]
\begin{center}
\includegraphics[width=0.51\textwidth]{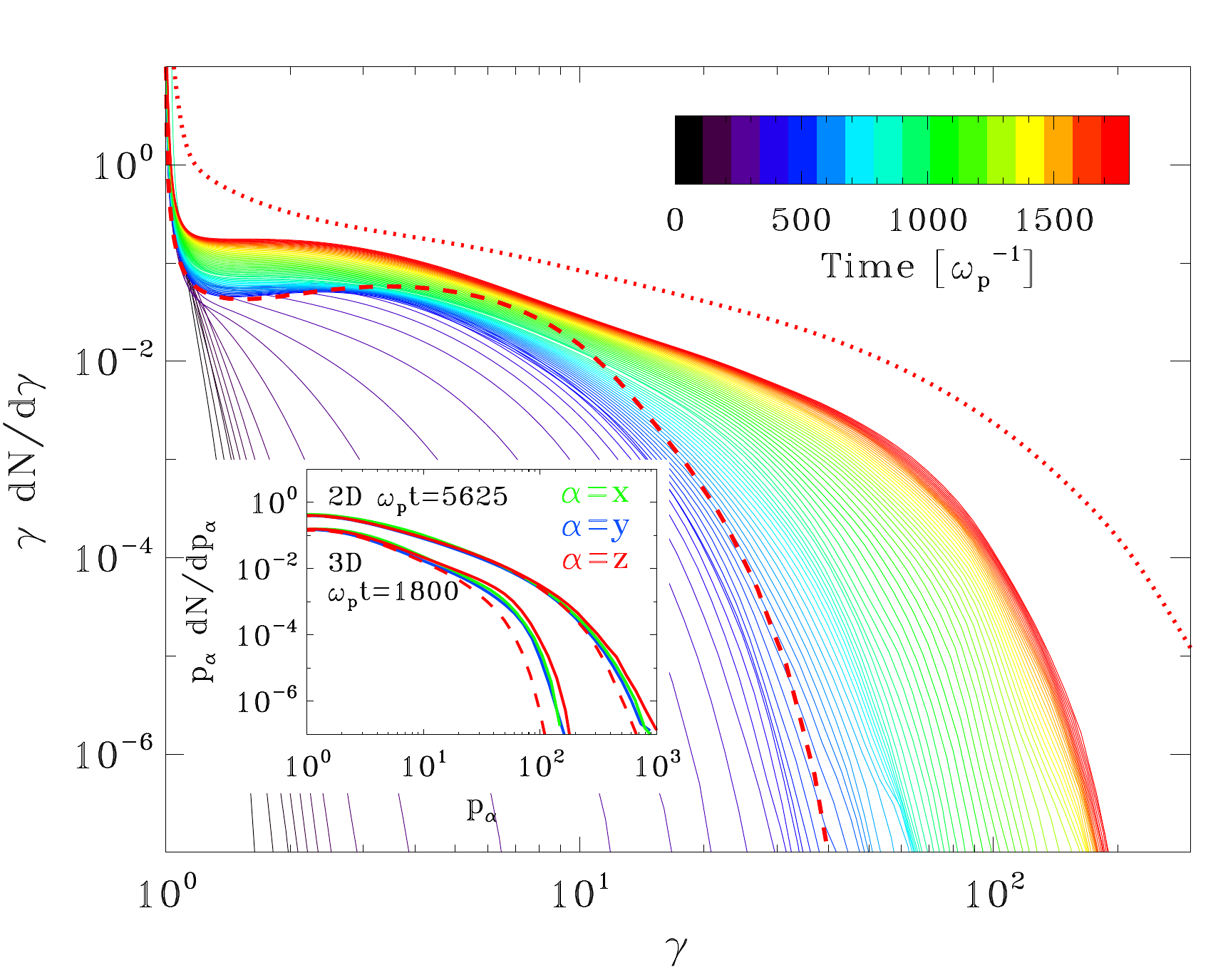}
\caption{Temporal evolution of particle energy spectrum, from a 3D simulation of $\sigma=10$ reconnection. The spectra from two 2D simulations with in-plane (out-of-plane, respectively) anti-parallel fields are shown with red dotted (dashed, respectively) lines. In the inset, positron momentum spectrum along $x$ (green), $y$ (blue), $+z$ (red solid) and $-z$ (red dashed), for 2D and 3D, as indicated.}
\label{fig:spec3d}
\end{center}
\end{figure}

\section{Spectrum and Acceleration Mechanism}\label{sec:accel}
In \fig{spec2d} we present the time evolution of the particle energy spectrum integrated over the whole reconnection region (more precisely, for $|y|\lesssim 500\comp$), from a 2D simulation with $\sigma=10$.\footnote{In our spectra, we do not include the hot particles that were initialized in the sheet to provide the pressure support against the external magnetic field. With this choice, the late-time spectrum is nearly independent from the current sheet initialization (S14).} At the X-lines, more than half of the initial magnetic energy is converted into particle kinetic energy. \fig{spec2d} shows that a self-consistent by-product of relativistic reconnection is the generation of a broad non-thermal spectrum extending to ultra-relativistic energies. For $\sigma=10$, the spectrum at $\gamma\gtrsim 1.5$ can be fitted with a power-law of slope $p\equiv - d\log N/d\log \gamma\sim2$ (dotted red line).\footnote{The peak at $\gamma\lesssim1.5$ contains the cold particles that are drifting towards the sheet at the reconnection speed $v_{\rm rec}\simeq0.08\,c$.} The spectrum clearly departs from a Maxwellian with mean energy $(\sigma+1)\,mc^2$  (red dashed line, assuming complete field dissipation). As shown in the inset of \fig{spec2d}, the power-law slope depends on the magnetization, being harder for higher $\sigma$ ($p\sim1.5$ for $\sigma=50$, compare solid and dotted green lines). The slope is steeper for lower magnetizations  ($p\sim4$ for $\sigma=1$, solid and dotted black lines), approaching the result from earlier studies of non-relativistic reconnection, that found poor acceleration efficiencies \citep[][]{drake_10}.

As described below, the power-law shape of the energy spectrum is established as the particles interact with the X-points, where they get accelerated by the reconnection electric field. After being advected into the major islands shown in \fig{fluid2d}a, the particles experience a variety of other acceleration processes \citep[][]{drake_06,oka_10}, yet the power-law index does not appreciably change. As described in S14, the anti-reconnection electric field between two merging islands plays a major role for the increase in the spectral cutoff shown in \fig{spec2d}. For magnetizations $\sigma\gtrsim10$ that yield $p\lesssim2$, the increase in maximum energy is expected to terminate, since the mean energy per particle cannot exceed $(\sigma+1)\,mc^2$.\footnote{For $\sigma\lesssim 10$ (so, $p\gtrsim 2$), the increase in maximum energy does not stop, but it slows down at late times. As the islands grow bigger they become slower, so the anti-reconnection electric field during mergers gets weaker.} For a power-law of index $1<p<2$ starting from $\gamma_{\rm min}=1$, the maximum Lorentz factor should saturate at $\gamma_{\rm max}\sim[(\sigma+1)(2-p)/(p-1)]^{1/(2-p)}$.

So far, we have shown that 2D simulations of relativistic reconnection produce hard populations of non-thermal particles. The validity of our conclusions may be questioned if the structure of X-points in 3D is significantly different from 2D. In particular, the DK mode is expected to result in heating, not in particle acceleration \citep{zenitani_07}. In \fig{spec3d} we follow the temporal evolution of the particle spectrum in a 3D simulation with $\sigma=10$. We confirm the conclusions of earlier studies \citep{zenitani_08,cerutti_13b}, that the spectrum at early times is quasi-thermal (black to cyan lines in \fig{spec3d}), and it resembles the distribution resulting from the DK mode (the red dashed line shows the spectrum from a 2D simulation with out-of-plane anti-parallel fields, to isolate the contribution of the DK mode). As shown in \S\ref{sec:struct}, the DK mode is the fastest to grow, but the sheet evolution at late times is controlled by the tearing instability, in analogy to 2D simulations with in-plane fields. The X-points formed by the tearing mode can efficiently accelerate non-thermal particles, and the spectrum at late times (cyan to red lines in \fig{spec3d}) presents a pronounced high-energy power-law. The power-law slope is $p\sim2.3$, close to the $p\sim2$ index of 2D simulations with in-plane fields. With respect to the 2D spectrum (dotted red line in  \fig{spec3d}), the normalization and the upper energy cutoff of the 3D spectrum are smaller, due to the lower reconnection rate ($v_{\rm rec}\simeq 0.02\,c$ in 3D versus $v_{\rm rec}\simeq 0.08\,c$ in 2D), so that fewer particles enter the current sheet per unit time, where they get accelerated by a weaker  electric field $E_{\rm rec}\sim (v_{\rm rec}/c)\, B_0$. In analogy to 2D, we argue that the anti-reconnection electric field in between merging plasmoids drives the increase in the high-energy spectral cutoff shown in \fig{spec3d}.

\begin{figure}[tbp]
\begin{center}
\includegraphics[width=0.51\textwidth]{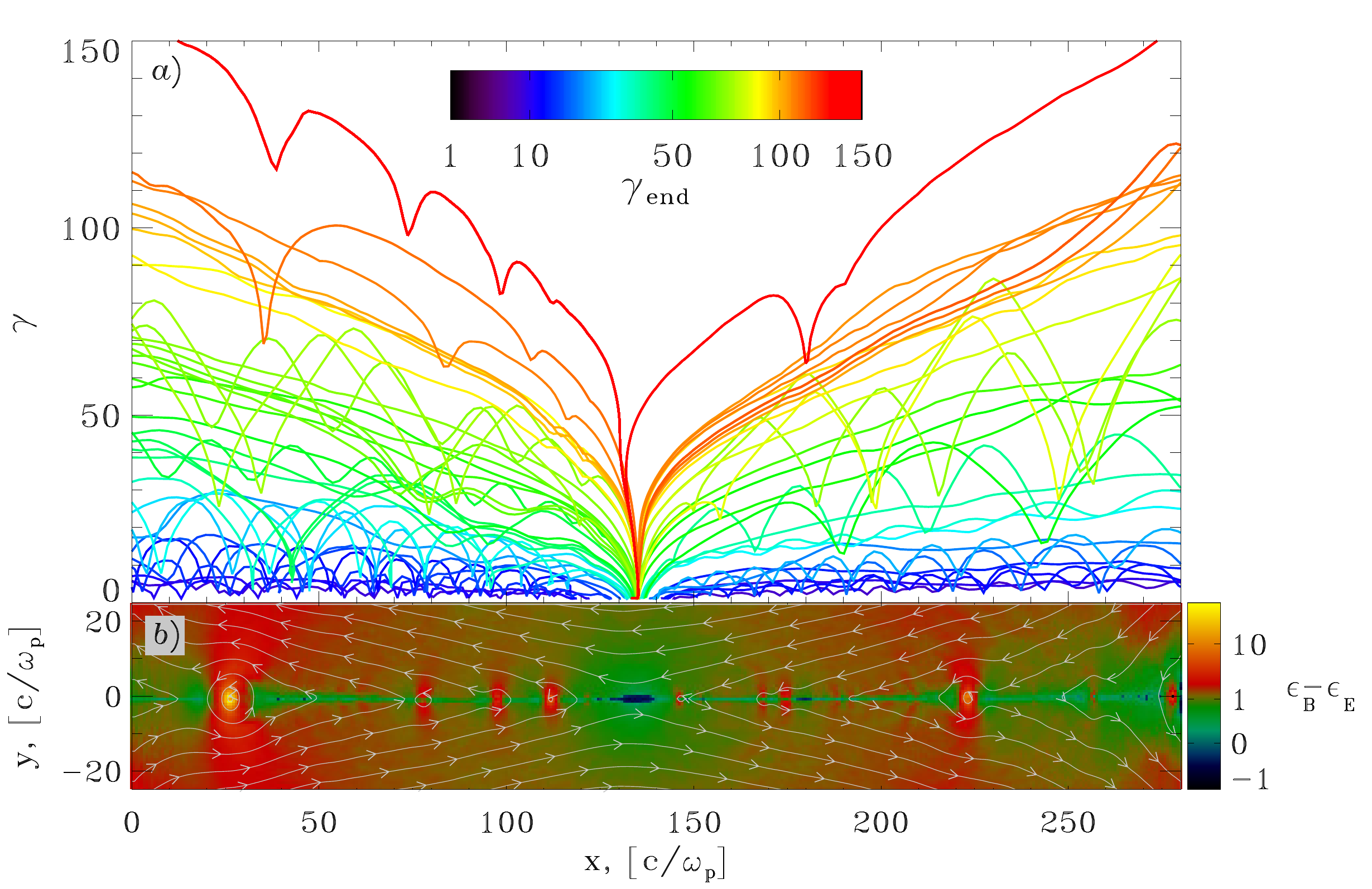}
\caption{(a) Energy evolution of a sample of selected particles interacting with a major X-point, as a function of the location $x$ along the current sheet. Colors are scaled with $\gamma_{\rm end}$, the  Lorentz factor attained at the boundary of the X-line (at $x=0$ or $280\comp$, depending on the particle). (b) $\epsilon_B-\epsilon_E$ at the time when the particles interact with the X-point (here, $\epsilon_E=E^2/8\pi m n c^2$).}
\label{fig:accel}
\end{center}
\end{figure}

The mechanism of particle acceleration at X-points is presented in \fig{accel}, in the case of a 2D simulation with $\sigma=10$. We follow the energy evolution of a sample of simulation particles that interact with the sheet in the vicinity of the central X-point (located at $x\sim 135\comp$). There, the magnetic energy is smaller than the electric energy (blue region in \fig{accel}b), so the  particles become unmagnetized and they get accelerated along $z$ by the reconnection electric field. The final energy of the particles -- the color in \fig{accel}a indicates the Lorentz factor measured at the boundary of the X-line -- directly correlates with the location at the moment of interaction with the sheet \citep[][]{larrabee_03,bessho_12}. Particles interacting closer to the center of the X-point (darkest blue in \fig{accel}b) are less prone to be advected away along $x$ by the reconnected magnetic field, so they can stay longer in the acceleration region and reach higher Lorentz factors (orange and red lines in \fig{accel}a). A broad distribution is then established, as a result of the different energy histories of particles interacting at different distances from the X-point. The particle spectrum at the edge of the acceleration region (the blue area in \fig{accel}(b)) resembles the form $dN/d\gamma\propto \gamma^{-1/4} \exp[-(4\gamma/\sigma\beta_{\rm rec}^2\lambda)^{1/2}]$ predicted by \citet{bessho_12}, where $\beta_{\rm rec}=v_{\rm rec}/c$ and $\lambda$ is the length of the acceleration region in units of the Larmor radius of the heated particles (having $\langle\gamma\rangle\sim \sigma$). The spectrum of particles flowing into the major islands still bears memory of this scaling, but it becomes softer, due to the addition of low-energy particles injected at the secondary X-points (weaker than the major X-point in \fig{accel}) or accreted onto the outflowing secondary islands.

After being accelerated along $z$, the particles are advected along $x$ by the reconnected magnetic field, and they finally enter the major islands shown in \fig{fluid2d}a. In the islands, the accelerated particles gyrate in the strong fields shown in \fig{fluid2d}c, isotropizing their angular distribution (which was strongly beamed along $z$ close to the X-points, and along $x$ in the reconnection exhausts, see \citealt{cerutti_13a}). Since most of the particles at late times are contained in the major islands, it is not surprising that the long-term momentum spectra show little signs of anisotropy (see the inset in \fig{spec3d}). Even the residual difference between the momentum spectra along $+z$ and $-z$ (red solid and dashed lines, respectively) diminishes at later times (the 2D momentum spectra at $\ompt=1800$ were similar to the 3D results in the inset of \fig{spec3d}, showing that the anisotropy decays over time).\footnote{The particle angular distribution at late times may be different in the presence of a guide field, see S14.}

\section{Discussion}
By means of large-scale PIC simulations, we have provided definitive evidence that non-thermal particle acceleration is a {\it generic} by-product of the long-term evolution of relativistic reconnection, in both 2D and 3D. We have focused on the case of anti-parallel reconnection without a guide field. We find that in 3D the DK mode delays the onset of efficient particle acceleration, but the physics at late times is similar to its 2D counterpart. Earlier studies could not capture the late-time evolution that leads to efficient particle acceleration, due to insufficient integration times and/or limited computational domains. The particles  accelerated by the reconnection electric field at the X-points form a power-law with slope $p$ that varies from $p\sim 4$ for $\sigma=1$ to $p\lesssim 1.5$ for $\sigma\gtrsim50$. For $\sigma\gtrsim10$ the index is $p\lesssim 2$, harder than in relativistic shocks \citep{sironi_spitkovsky_11a,sironi_13}. So, relativistic reconnection is a viable candidate for producing hard spectra in astrophysical non-thermal sources.

Flat electron spectra below GeV energies are required to fit the broad-band emission of hotspots in radio galaxies \citep{stawarz_07} and the X-ray spectrum of blazars \citep{celotti_08,sikora_09}. Also, relativistic reconnection in blazars might explain the recently discovered ultra-fast ($\sim3-5$ minutes) TeV flares \citep{aharonian_07,giannios_09,nalewajko_11}.
In PWNe, the recently detected GeV flares from the Crab Nebula require hard particle spectra with $p\lesssim 2$ \citep{buehler_12}, and relativistic reconnection has been invoked to explain the temporal variability of the flares \citep{cerutti_13a}.
Our first-principles simulations provide a physically-grounded model, based on relativistic reconnection, for the generation of hard particle spectra in astrophysics.

\acknowledgements
L.S. is supported by Einstein grant PF1-120090, A.S. by NASA grants NNX12AD01G and NNX13AO80G and Simons Foundation grant 267233. The simulations used PICSciE-OIT HPCC at Princeton University, XSEDE under contract TG-AST120010, and NASA HEC.


\vspace{0.3in}

\begin{thebibliography}{48}
\expandafter\ifx\csname natexlab\endcsname\relax\def\natexlab#1{#1}\fi

\bibitem[{{Aharonian} \& al.(2007)}]{aharonian_07}
{Aharonian}, F. \& al. 2007, \apjl, 664, L71

\bibitem[{{Bessho} \& {Bhattacharjee}(2005)}]{bessho_05}
{Bessho}, N. \& {Bhattacharjee}, A. 2005, Physical Review Letters, 95, 245001

\bibitem[{{Bessho} \& {Bhattacharjee}(2007)}]{bessho_07}
---. 2007, Physics of Plasmas, 14, 056503

\bibitem[{{Bessho} \& {Bhattacharjee}(2012)}]{bessho_12}
---. 2012, \apj, 750, 129

\bibitem[{{Buehler} \& al.(2012)}]{buehler_12}
{Buehler}, R. \& al. 2012, \apj, 749, 26

\bibitem[{{Buneman}(1993)}]{buneman_93}
{Buneman}, O. 1993, {in ``Computer Space Plasma Physics'', Terra Scientific,
  Tokyo, 67}

\bibitem[{{Celotti} \& {Ghisellini}(2008)}]{celotti_08}
{Celotti}, A. \& {Ghisellini}, G. 2008, \mnras, 385, 283

\bibitem[{{Cerutti} {et~al.}(2012){Cerutti}, {Uzdensky}, \&
  {Begelman}}]{cerutti_12a}
{Cerutti}, B., {Uzdensky}, D.~A., \& {Begelman}, M.~C. 2012, \apj, 746, 148

\bibitem[{{Cerutti} {et~al.}(2013{\natexlab{a}}){Cerutti}, {Werner},
  {Uzdensky}, \& {Begelman}}]{cerutti_13a}
{Cerutti}, B., {Werner}, G.~R., {Uzdensky}, D.~A., \& {Begelman}, M.~C.
  2013{\natexlab{a}}, \apj, 770, 147

\bibitem[{{Cerutti} {et~al.}(2013{\natexlab{b}}){Cerutti}, {Werner},
  {Uzdensky}, \& {Begelman}}]{cerutti_13b}
---. 2013{\natexlab{b}}, ArXiv:astro-ph/1311.2605

\bibitem[{{Daughton} \& {Karimabadi}(2007)}]{daughton_07}
{Daughton}, W. \& {Karimabadi}, H. 2007, Physics of Plasmas, 14, 072303

\bibitem[{{Drake} {et~al.}(2010){Drake}, {Opher}, {Swisdak}, \&
  {Chamoun}}]{drake_10}
{Drake}, J.~F., {Opher}, M., {Swisdak}, M., \& {Chamoun}, J.~N. 2010, \apj,
  709, 963

\bibitem[{{Drake} {et~al.}(2006){Drake}, {Swisdak}, {Che}, \&
  {Shay}}]{drake_06}
{Drake}, J.~F., {Swisdak}, M., {Che}, H., \& {Shay}, M.~A. 2006, \nat, 443, 553

\bibitem[{{Giannios}(2008)}]{giannios_08}
{Giannios}, D. 2008, \aap, 480, 305

\bibitem[{{Giannios}(2013)}]{giannios_13}
---. 2013, \mnras, 431, 355

\bibitem[{{Giannios} {et~al.}(2009){Giannios}, {Uzdensky}, \&
  {Begelman}}]{giannios_09}
{Giannios}, D., {Uzdensky}, D.~A., \& {Begelman}, M.~C. 2009, \mnras, 395, L29

\bibitem[{{Hoshino} \& {Lyubarsky}(2012)}]{hoshino_lyubarsky_12}
{Hoshino}, M. \& {Lyubarsky}, Y. 2012, \ssr, 173, 521

\bibitem[{{Jaroschek} {et~al.}(2004){Jaroschek}, {Lesch}, \&
  {Treumann}}]{jaroschek_04}
{Jaroschek}, C.~H., {Lesch}, H., \& {Treumann}, R.~A. 2004, \apjl, 605, L9

\bibitem[{{Kagan} {et~al.}(2013){Kagan}, {Milosavljevi{\'c}}, \&
  {Spitkovsky}}]{kagan_13}
{Kagan}, D., {Milosavljevi{\'c}}, M., \& {Spitkovsky}, A. 2013, \apj, 774, 41

\bibitem[{{Kirk} \& {Skj{\ae}raasen}(2003)}]{kirk_sk_03}
{Kirk}, J.~G. \& {Skj{\ae}raasen}, O. 2003, \apj, 591, 366

\bibitem[{{Larrabee} {et~al.}(2003){Larrabee}, {Lovelace}, \&
  {Romanova}}]{larrabee_03}
{Larrabee}, D.~A., {Lovelace}, R.~V.~E., \& {Romanova}, M.~M. 2003, \apj, 586,
  72

\bibitem[{{Liu} {et~al.}(2011){Liu}, {Li}, {Yin}, {Albright}, {Bowers}, \&
  {Liang}}]{liu_11}
{Liu}, W., {Li}, H., {Yin}, L., {Albright}, B.~J., {Bowers}, K.~J., \& {Liang},
  E.~P. 2011, Physics of Plasmas, 18, 052105

\bibitem[{{Lyubarsky} \& {Kirk}(2001)}]{lyubarsky_kirk_01}
{Lyubarsky}, Y. \& {Kirk}, J.~G. 2001, \apj, 547, 437

\bibitem[{{Lyubarsky} \& {Liverts}(2008)}]{lyubarsky_liverts_08}
{Lyubarsky}, Y. \& {Liverts}, M. 2008, \apj, 682, 1436

\bibitem[{{Lyubarsky}(2003)}]{lyubarsky_03}
{Lyubarsky}, Y.~E. 2003, \mnras, 345, 153

\bibitem[{{Lyubarsky}(2005)}]{lyubarsky_05}
---. 2005, \mnras, 358, 113

\bibitem[{{Lyutikov} \& {Blandford}(2003)}]{lyutikov_03}
{Lyutikov}, M. \& {Blandford}, R. 2003, ArXiv:astro-ph/0312347

\bibitem[{{Lyutikov} \& {Uzdensky}(2003)}]{lyutikov_uzdensky_03}
{Lyutikov}, M. \& {Uzdensky}, D. 2003, \apj, 589, 893

\bibitem[{{Nalewajko} {et~al.}(2011){Nalewajko}, {Giannios}, {Begelman},
  {Uzdensky}, \& {Sikora}}]{nalewajko_11}
{Nalewajko}, K., {Giannios}, D., {Begelman}, M.~C., {Uzdensky}, D.~A., \&
  {Sikora}, M. 2011, \mnras, 413, 333

\bibitem[{{Oka} {et~al.}(2010){Oka}, {Phan}, {Krucker}, {Fujimoto}, \&
  {Shinohara}}]{oka_10}
{Oka}, M., {Phan}, T., {Krucker}, S., {Fujimoto}, M., \& {Shinohara}, I. 2010,
  \apj, 714, 915

\bibitem[{{P{\'e}tri} \& {Lyubarsky}(2007)}]{petri_lyubarsky_07}
{P{\'e}tri}, J. \& {Lyubarsky}, Y. 2007, \aap, 473, 683

\bibitem[{{Romanova} \& {Lovelace}(1992)}]{romanova_92}
{Romanova}, M.~M. \& {Lovelace}, R.~V.~E. 1992, \aap, 262, 26

\bibitem[{{Sikora} {et~al.}(2009){Sikora}, {Stawarz}, {Moderski}, {Nalewajko},
  \& {Madejski}}]{sikora_09}
{Sikora}, M., {Stawarz}, {\L}., {Moderski}, R., {Nalewajko}, K., \& {Madejski},
  G.~M. 2009, \apj, 704, 38

\bibitem[{{Sironi} \& {Spitkovsky}(2011{\natexlab{a}})}]{sironi_spitkovsky_11b}
{Sironi}, L. \& {Spitkovsky}, A. 2011{\natexlab{a}}, \apj, 741, 39

\bibitem[{{Sironi} \& {Spitkovsky}(2011{\natexlab{b}})}]{sironi_spitkovsky_11a}
---. 2011{\natexlab{b}}, \apj, 726, 75

\bibitem[{{Sironi} \& {Spitkovsky}(2012)}]{sironi_spitkovsky_12}
---. 2012, Computational Science and Discovery, 5, 014014

\bibitem[{{Sironi} {et~al.}(2013){Sironi}, {Spitkovsky}, \&
  {Arons}}]{sironi_13}
{Sironi}, L., {Spitkovsky}, A., \& {Arons}, J. 2013, \apj, 771, 54

\bibitem[{{Spitkovsky}(2005)}]{spitkovsky_05}
{Spitkovsky}, A. 2005, in AIP Conf. Ser., Vol. 801, Astrophysical Sources of
  High Energy Particles and Radiation, ed. {T.~Bulik, B.~Rudak, \&
  G.~Madejski}, 345

\bibitem[{{Spruit}(2010)}]{spruit_10}
{Spruit}, H.~C. 2010, in Lecture Notes in Physics, Berlin Springer Verlag, Vol.
  794, Lecture Notes in Physics, Berlin Springer Verlag, ed. T.~{Belloni}, 233

\bibitem[{{Spruit} {et~al.}(2001){Spruit}, {Daigne}, \&
  {Drenkhahn}}]{spruit_01}
{Spruit}, H.~C., {Daigne}, F., \& {Drenkhahn}, G. 2001, \aap, 369, 694

\bibitem[{{Stawarz} {et~al.}(2007){Stawarz}, {Cheung}, {Harris}, \&
  {Ostrowski}}]{stawarz_07}
{Stawarz}, {\L}., {Cheung}, C.~C., {Harris}, D.~E., \& {Ostrowski}, M. 2007,
  \apj, 662, 213

\bibitem[{{Thompson}(1994)}]{thompson_94}
{Thompson}, C. 1994, \mnras, 270, 480

\bibitem[{{Thompson}(2006)}]{thompson_06}
---. 2006, \apj, 651, 333

\bibitem[{{Uzdensky} {et~al.}(2010){Uzdensky}, {Loureiro}, \&
  {Schekochihin}}]{uzdensky_10}
{Uzdensky}, D.~A., {Loureiro}, N.~F., \& {Schekochihin}, A.~A. 2010, Physical
  Review Letters, 105, 235002

\bibitem[{{Yin} {et~al.}(2008){Yin}, {Daughton}, {Karimabadi}, {Albright},
  {Bowers}, \& {Margulies}}]{yin_08}
{Yin}, L., {Daughton}, W., {Karimabadi}, H., {Albright}, B.~J., {Bowers},
  K.~J., \& {Margulies}, J. 2008, Physical Review Letters, 101, 125001

\bibitem[{{Zenitani} \& {Hoshino}(2001)}]{zenitani_01}
{Zenitani}, S. \& {Hoshino}, M. 2001, \apjl, 562, L63

\bibitem[{{Zenitani} \& {Hoshino}(2007)}]{zenitani_07}
---. 2007, \apj, 670, 702

\bibitem[{{Zenitani} \& {Hoshino}(2008)}]{zenitani_08}
---. 2008, \apj, 677, 530

\end{thebibliography}

\end{document}